\documentclass[a4paper,11pt]{article}            
\usepackage{cite}
\usepackage{graphicx}
\usepackage{float}
\usepackage{wrapfig}
\usepackage{authblk}
\begin{document}
\title{Revisiting the Coulomb-Damped Harmonic Oscillator}
\author{Joseph A Rizcallah\\joeriz68@gmail.com}
\affil{School of Education, Lebanese University, Beirut, Lebanon}
\date{}
\maketitle

\begin{abstract}
\noindent
The force of dry friction is studied extensively in introductory physics but its effect on oscillations is hardly ever mentioned. Instead, to provide a mathematically tractable introduction to damping, virtually all authors adopt a viscous resistive force. While exposure to linear damping is of paramount importance to the student of physics, the omission of Coulomb damping might have a negative impact on the way the students conceive of the subject. In the paper, we propose to approximate the action of Coulomb friction on a harmonic oscillator by a sinusoidal resistive force whose amplitude is the model's only free parameter. We seek the value of this parameter that yields the best fit and obtain a closed-form analytic solution, which is shown to nicely fit the numerical one.
\end{abstract}

\section{Introduction}
Introductory-level textbooks~\cite{yf,tm,fr,kg} discuss damped oscillations predominantly within the linear resistive force model. This presents a fairly simple and familiar mathematical problem that can be solved exactly using standard techniques in ordinary differential equations while at the same time serving a good introduction to the basics of damping. Moreover, the study of linear damping is of great importance, as it can be used to model damping in a viscous medium, magnetic damping, the damping of electromagnetic oscillations in RLC circuits, as well as the so called wave radiation damping. In practice, however, the linear model has its limitations. For one, it fails to describe the damping due to sliding friction~\cite{simpri,zon,minsik}, often called Coulomb damping, which is omitted in most introductory treatments due presumably to the piecewise nature of the exact solution~\cite{simpri,mol}.

This state of affairs presents a pedagogical challenge for the physics teachers. On the one hand, in the name of mathematical simplicity, they are willing to accept the independence of the force of dry friction on the speed throughout mechanics, but on the other they quickly abandon it in favour of the linear model, when it comes to oscillations. Some authors seem to have noted this inconsistency in the past (see for example~\cite{clifmin}) and several suggestions have been made in the literature to address Coulomb damping at the introductory level. However, most of these attempts were limited to analyzing the amplitude and energy decay over half-cycles~\cite{mol,marabb,kam,lap} or had to make some kind of approximation, oftentimes involving energy, to derive an approximate closed-form analytic solution~\cite{wansch,squ}. Recently, a Fourier series approach has been employed in~\cite{vit} to obtain both exact and approximate analytic solutions to the combined Coulomb and linearly damped oscillator. A more recent discussion of the combined effect of viscous and sliding friction on damping can be found in~\cite{hin}.    

In the paper, after a brief overview of Coulomb damping, we propose a resolution to this pedagogical difficulty based on approximating the force of Coulomb friction by an appropriate sinusoidal resistive force. Our approach involves elements of inhomogeneous linear differential equations of the 2nd order and should be accessible to students familiar with the elementary mathematical description of linear damping, forced oscillations, and resonance. While our treatment here leads to the same approximate solution obtained in \cite{vit} (in the absence of linear damping), it unfolds along totally different lines, keeping the exposition accessible to as wide an audience as possible.

\section{Overview of Coulomb damping}
In this section we set up the problem of Coulomb damping and recall some of the properties of its exact solution. To this end, consider a spring-mass system moving over a rough horizontal surface. Designate by $m$ and $k$ respectively the mass attached to the massless spring and the stiffness of the latter. Let $f$ denote the magnitude of the force of sliding friction that acts against the motion. To be sure, we are interested in the case of weak friction or underdamping, whereby the system performs many oscillations before coming to rest. Throughout the paper we assume that the system starts from rest with a sufficiently large $x(0) = x_0 > 0$. The motion of the spring-mass system is governed by the differential equation
\begin{equation}
\label{mainode}
m\ddot{x}+kx=\pm f	
\end{equation}
where $x(t)$ is the elongation of the spring, the overdot denotes differentiation w.r.t. to time $t$, and the $+/-$ sign in the right-hand side corresponds to motion in the negative/positive sense of the $x$-axis respectively, i.e. to negative/positive velocity $\dot{x}$. Note that changes in the direction of the friction force take place at the turning points and cause jump discontinuities of magnitude $2f/m$ in the acceleration.

\begin{figure}[H]
\centering
\includegraphics[scale=1]{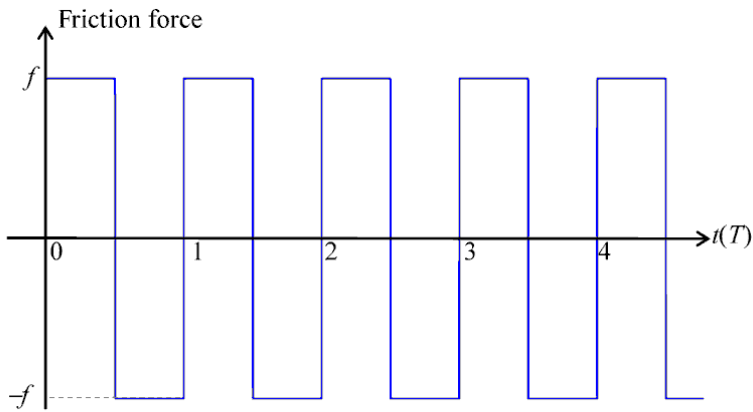}
\caption{The time graph of the force of sliding friction displaying periodic alternation between $f$ and $-f$.}
\label{fig1}
\end{figure}

Introducing the new variable $X=x \mp f/k$, we see from (\ref{mainode}) that between any two consecutive turning points the motion is simple harmonic, of period $T = 2\pi\sqrt{m/k}$, about an equilibrium position displaced by a fixed amount $f/k$ against the motion (see e.g. \cite{barr}). It follows that the time it takes to go from one turning point to the next (half a cycle) is a constant equal to half the period of the undamped oscillator. Thus, as in the case of linear damping, the motion is quasiperiodic, but unlike the latter the quasiperiod is equal to the period of the undamped oscillator. An important consequence of this quasiperiodicity is the “periodicity” of the force of sliding friction which, as illustrated in figure~\ref{fig1}, is seen to flip-flop between $f$ and $-f$ as long as the mass is moving.

The damping effect due to sliding friction arises from the cyclic offset in the equilibrium position against the motion. To see this, suffice it to consider the motion separately over even and odd half cycles. Let $x_n$ denote the turning point at the end on the $n$th half cycle. For $x_n > 0$ (even $n$), the equilibrium position is shifted by $+f/k$ during the $n + 1$st half cycle, the effective amplitude is $x_n-f/k$ and so the mass stops at $x_{n+1} = f/k - (x_n - f/k) = - x_n + 2f/k$. Similarly, for $x_n < 0$ (odd $n$), one has an offset of $-f/k$ and an effective amplitude of $-f/k - x_n$ yielding a turning point at $x_{n+1} = -f/k + (-f/k -x_n) = - x_n - 2f/k$. Reasoning by mathematical induction, we conclude that the turning points occur at
\begin{equation}
\label{turnpts}
x_n=(-1)^n(x_0-2nf/k)	
\end{equation} 
at the end of the $n$th half cycle. Note that equation (\ref{turnpts}) is physically meaningful as long as the expression in parentheses is nonnegative, i.e. for $n < kx_0/2f$. This allows one to quantitatively define underdamping by the inequality $kx_0/f >> 1$. Finally, from (\ref{turnpts}) we see that the distance between the turning points is contracting by the fixed amount $2f/k$ every half cycle, a manifestation of linear amplitude decay which is a well-known feature of Coulomb damping.

\section{Sinusoidal resistive force}
We now intend to modify equation (\ref{mainode}) in such a way as to be able to find a simple closed-form approximate solution. The periodicity of the force of friction suggests replacing the right-hand side of (\ref{mainode}) by a sinusoidal resistive force of angular frequency $\omega = \sqrt{k/m}$ and unknown amplitude $F$. We are thus led to consider
\begin{equation}
\label{appode}
\ddot{x}+\omega^2 x=(F/m)\sin{\omega t}
\end{equation} 
In fact, the right-hand side of (\ref{appode}) is the first harmonic in the Fourier series expansion (in odd harmonics) of the periodic force of sliding friction, and its amplitude $F$ can be easily found as one of the Fourier coefficients (for more details, see~\cite{vit}). However, the Fourier series approach is deemed to be beyond the scope of the intended audience and the next section is devoted to determining the value of $F$ that yields the best agreement with the exact solution, without recourse to Fourier series. In the meantime, we shall assume $F$ to be small enough for underdamped oscillations to occur, and turn our attention to the solution of the initial value problem for (\ref{appode}), subject to the initial conditions $x(0) = x_0$ and $\dot{x}(0)=0$. To this end, let us consider the equation of forced oscillations $\ddot{x}+\omega^2 x=(F/m)\sin{\Omega t}$, which goes over to (\ref{appode}) at resonance $\Omega = \omega$. From the elementary theory of linear differential equations it is well-known that the solution of this equation is given by
$$x(t)= A\cos{\omega t} + B\sin{\omega t} - \frac{F/m}{\Omega^2-\omega^2}\sin{\Omega t},$$ 
where the constants $A = x_0$ and $B=\frac{F/m}{\Omega^2-\omega^2}\frac{\Omega}{\omega}$ are determined using the initial conditions. Substituting these in the above expression and passing to the limit $\Omega \rightarrow \omega$ we arrive at the solution
\begin{equation}
\label{appsol}
x(t) = \left(x_0-\frac{F}{2k}{\omega t}\right)\cos{\omega t} +\frac{F}{2k}\sin{\omega t} 
\end{equation} 
In the weak-friction regime, the first term in (\ref{appsol}) can be interpreted as a quasiperiodic oscillation with a linearly decaying amplitude, as illustrated in figure~\ref{fig2} below. It is worth noting that the decay here arises as a result of resonance between the sinusoidal resistive force and the spring-mass system. Needless to say, the buildup of oscillation seen in figure 2 is an unphysical artifact caused by extrapolating the solution past the halting point. The last term in (\ref{appsol}) does not seem to contribute significantly until the motion is about to cease, i.e. when $x_0-\frac{F}{2k}{\omega t} = \frac{F}{2k}$. Its presence, however, is essential to meet the initial conditions as well as for the approximate and exact solutions to attain their turning points simultaneously (see below). 
\begin{figure}[H]
\centering
\includegraphics[scale=1]{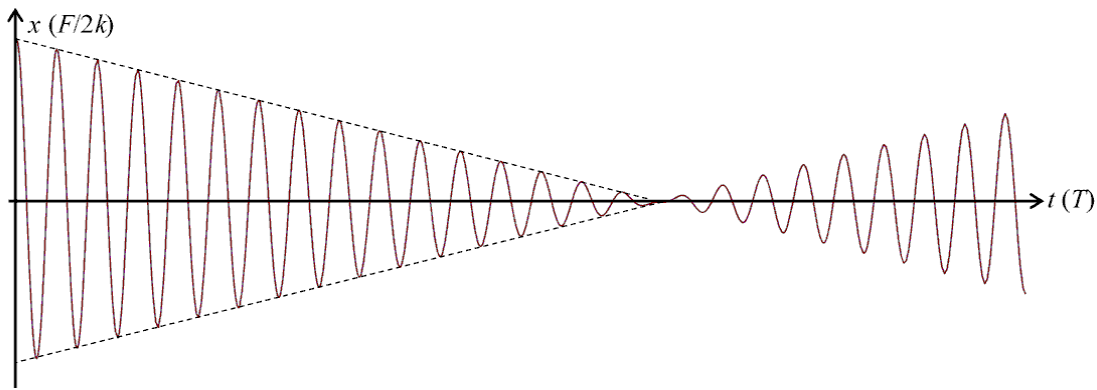}
\caption{The damped oscillations of a spring-mass system with a sinusoidal resistive force exhibit the characteristic linear decay of Coulomb damping.}
\label{fig2}
\end{figure}

\section{Determining the amplitude $F$}
We now turn to the amplitude $F$ of the sinusoidal resistive force. Our goal is to determine the value of $F$ that ensures the best agreement between the approximate and exact solutions. From a student's perspective, the following three conjectures seem plausible: (1) reasoning by analogy with ac voltage/current one may surmise that the sinusoidal resistive force and sliding friction should have equal rms values, or (2) guided by Newton's 2nd law one may contemplate the equality of impulses (or time averages) exerted by the forces over every half cycle, and last but not least, (3) on purely mathematical grounds one may try and minimize the mean square difference between these forces. As we show below, option (3) yields the best fit, but in the absence of an explicit criterion, it is not yet clear what is meant by “best”. This will be dealt with shortly. Presently, we would like to pursue our guesses.

\begin{wrapfigure}{r}{0.5\textwidth}
\begin{center}
\includegraphics[scale=0.5]{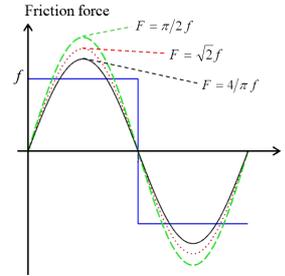}
\end{center}
\caption{The graphical representations of the three sinusoidal resistive forces discussed in the text displayed against that of friction.}
\label{fig3}
\end{wrapfigure}

Let us begin with the equality of the rms values: $\int^{T}_{0}{F^2 \sin^2{\omega t} dt}=\int^{T}_{0}{f^2dt}$. This implies $F=\sqrt{2}f$. On the other hand, the equality of impulses $\int^{T/2}_{0}{F\sin{\omega t} dt}=\int^{T/2}_{0}{f dt}$ leads to $F=(\pi/2)f$, while the least mean square difference requires minimizing the integral $\int^{T/2}_{0}{(f-F \sin{\omega t})^2 dt}$ w.r.t. the parameter $F$, which yields the equation $\int^{T/2}_{0}{(f-F \sin{\omega t})\sin{\omega t} dt}=0$ whose solution is readily found to be $F=(4/\pi)f$. Figure~\ref{fig3} shows the time graphs, over a period, of the three sinusoidal forces corresponding to those amplitudes along with the force of sliding fiction.  

\begin{figure}[H]
\centering
\includegraphics[scale=1]{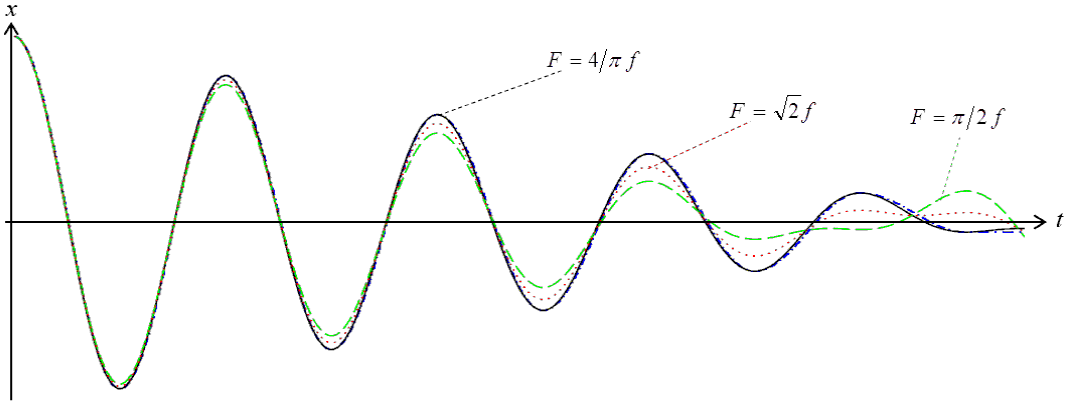}
\caption{The position-time graphs of a numerical solution of equation (1) and the corresponding analytic solutions (4) associated with the three amplitudes discussed in the text. The color coding and line style match those of figure~\ref{fig3}.}\label{fig4}
\end{figure}

How can we tell which of these (or other) sine curves best fit the square signal? One way to go about answering this question is to find the solution corresponding to each and compare it with the exact one. This, however, requires the nontrivial task of defining and minimizing the “distance” between two functions (at a continuum of points), which we believe is beyond the scope of the targeted audience. Instead, we suggest to compare the solutions only at a finite number of points, namely the turning points. Thus, the goodness of fit of an approximate solution (4) is judged by how close its turning points come to those of the exact solution (\ref{turnpts}). More precisely, differentiating (\ref{appsol}) w.r.t. time and equating the result to zero we find, regardless of the value of $F$, that the turning points occur at $t_n=nT/2$, which upon substitution back in (\ref{appsol}), and after some simplification, yields
$$x_n=(-1)^n\left(x_0-n\frac{\pi}{2}\frac{F}{k}\right).$$
We here have ignored the additional turning point that may arise from the decaying amplitude. Excepting the latter, all the turning points of the approximate solutions are attained at the right times, i.e. simultaneously with the exact one. We therefore seek the amplitude $F$ that makes the $x$-coordinates of these points agree with those of (\ref{turnpts}) as much as possible. On comparison with (\ref{turnpts}), one readily sees that “perfect” agreement is achieved for $F=(4/\pi)f$. Therefore, in the class of periodic resistive forces the “least mean square” approximation yields the best fit. (As noted earlier, this is the amplitude of the first harmonic in the Fourier series expansion of the periodic force of sliding friction (see~\cite{vit})). Figure~\ref{fig4} depicts a numerical solution of (\ref{mainode}) (obtained in Maple) together with the corresponding analytic solutions (\ref{appsol}) associated with the above three approximations. It is clear that both the "rms" and "impulse" approximations produce too fast a decay. By contrast, as it appears from the figure, the "least mean square" solution (solid black) so closely matches the numerical one (dash-dotted blue), almost everywhere, that one can hardly make out their graphs; only near the halting point do the graphs begin to somewhat diverge.

To sum it all up, the analytic solution of (\ref{appode}) that best fits the exact underdamped solution of (\ref{mainode}) is given by
\begin{equation}
\label{bestfit}
x(t) = \left(x_0-\frac{2f}{\pi k}{\omega t}\right)\cos{\omega t} +\frac{2f}{\pi k}\sin{\omega t}.	
\end{equation}
In the next section we employ this fairly simple expression to explore the energy decay for the Coulomb damped oscillator.

\section{Energy decay}
The negative work done by friction on the spring-mass system has the effect of dissipating its energy $E$.
Using the formula $E = \frac{1}{2}kx^2 + \frac{1}{2}m\dot{x}^2$ together with (\ref{bestfit}) and its time derivative $\dot{x}(t)=-\omega\left(x_0-\frac{2f}{\pi k}{\omega t}\right)\sin{\omega t}$, we get
$$E=E_0\left[\left(1-\frac{2f}{\pi kx_0}{\omega t}\right)^2+\left(\frac{2f}{\pi kx_0}\right)^2\sin^2{\omega t}+\left(1-\frac{2f}{\pi kx_0}{\omega t}\right)\left(\frac{2f}{\pi kx_0}\right)\sin{\omega t}\right]$$
where $E_0=1/2kx_{0}^2$ is the system's initial energy. In the weak-friction regime $f/kx_0 <<1$, the energy decay is dominated by the first term in the bracket on the right-hand side, while the two oscillating terms cause little ripples around the decay trend (for further details, see~\cite{kar,pal}). Ignoring these ripples, after some manipulation, we find
\begin{equation}
\label{nrgtrnd}
E=E_0\left(1-\frac{2f}{\pi\sqrt{2mE_0}}t\right)^2,
\end{equation}
which indicates a parabolic decay trend. Figure 5 depicts the energy step-like decay according to the exact solution, the best-fit approximate solution (\ref{bestfit}), as well as the parabolic trend line (\ref{nrgtrnd}). Again note the excellent agreement between the exact and approximate decay curves, particularly at the turning points where the two flatten out. 

\begin{wrapfigure}{r}{0.45\textwidth}
\begin{center}
\includegraphics[scale=0.7]{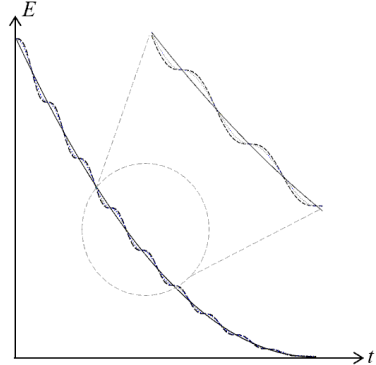}
\end{center}
\caption{The energy decay curves of the Coulomb underdamped oscillator predicted by the exact solution (dotted blue) and the approximate solution (\ref{bestfit}) (dashed black) along with the trend line (\ref{nrgtrnd}) (solid).}
\label{fig5}
\end{wrapfigure}
It is instructive to compare (\ref{nrgtrnd}) with the energy of a particle of mass $m$ acted upon by a constant force of friction $f$. For the latter, one has $E=E_0\left(1-\frac{f}{\sqrt{2mE_0}}t\right)^2$ (for the kinetic energy) which differs from (\ref{nrgtrnd}) by the dimensionless factor $2/\pi$. Being less than unity, this extra factor makes the energy decay rate slower for the underdamped oscillator. The difference in the decay rates is a direct consequence of the fact that only the kinetic energy is subject, so to speak, to the eroding action of friction. Therefore, the particle's energy, being purely kinetic, is eroded by friction at a higher rate than the oscillator's energy which is partly "hidden away" from friction in potential form. Note that the same is true for linear underdamping, where the (exponential) decay of the oscillator's energy takes place at half as fast a rate as that of a particle subject to the same friction force.

\section{Discussion}
We put forward a fairly simple model of the motion of the Coulomb damped harmonic oscillator based on approximating the damping action of Coulomb friction by a sinusoidal resistive force whose amplitude $F$ is the model's sole parameter. In effect, the model is a first approximation to that obtained from a full Fourier series expansion of the periodic force of sliding friction (see~\cite{vit}). However, in contrast to~\cite{vit}, where the solution follows automatically from the theory of Fourier series, we develop an elementary approach which leads in a simple and explicit manner to the same approximate solution. In particular, we discuss some plausible values of the model's parameter and suggest a method to determine the one that best fits the exact solution. The thus obtained analytic solution is shown to agree well with the numerical one and is employed to study the energy decay of the Coulomb underdamped oscillator.  

It should come as no surprise to learn that the suggested approach has its limitations. For example, it is not clear how (\ref{bestfit}) can be used to predict the halting point, which according to the exact solution occurs at a turning point with $|x_n|\leq f/k$ (we make no distinction between static and kinetic friction), or equivalently $kx_0/2f -1/2 \leq n \leq kx_0/2f + 1/2$. Moreover, being a smooth function of time, the approximate solution has a continuous acceleration, in contrast with the exact solution, whose acceleration, you recall, admits jump discontinuities at the turning points. In this regard, it is interesting to note that, at the turning points, the approximate solution's acceleration $-\omega^2x_n$ is halfway between the left and right limits $-\omega^2x_n \pm f/m$ of the exact one. (A similar issue arises in the Fourier series approach in the form of the Gibbs phenomenon.)  

Finally, we believe that our approach, apart from providing the students and their instructors with a simple way to analyze the motion of the Coulomb damped harmonic oscillator, offers a fresh insight into the origin of linear amplitude drop off and the associated quadratic energy decay. We hope that our method will make the Coulomb damping problem more accessible to anyone familiar with the basics of forced oscillations and resonance, thus help bridge a gap in the existing literature.

\end{document}